\documentclass[]{article}
\usepackage{epsfig}

\hoffset= - 1.0in         

\catcode`\@=11
\def\marginnote#1{}

\newcount\hour
\newcount\minute
\newtoks\amorpm
\hour=\time\divide\hour by60
\minute=\time{\multiply\hour by60 \global\advance\minute by-\hour}
\edef\standardtime{{\ifnum\hour<12 \global\amorpm={am}%
        \else\global\amorpm={pm}\advance\hour by-12 \fi
        \ifnum\hour=0 \hour=12 \fi
        \number\hour:\ifnum\minute<10 0\fi\number\minute\the\amorpm}}
\edef\militarytime{\number\hour:\ifnum\minute<10 0\fi\number\minute}

\def\draftlabel#1{{\@bsphack\if@filesw {\let\thepage\relax
   \xdef\@gtempa{\write\@auxout{\string
      \newlabel{#1}{{\@currentlabel}{\thepage}}}}}\@gtempa
   \if@nobreak \ifvmode\nobreak\fi\fi\fi\@esphack}
        \gdef\@eqnlabel{#1}}
\def\@eqnlabel{}
\def\@vacuum{}
\def\draftmarginnote#1{\marginpar{\raggedright\scriptsize\tt#1}}

\def\draft{\oddsidemargin -.5truein
        \def\@oddfoot{\sl preliminary draft \hfil
        \rm\thepage\hfil\sl\today\quad\militarytime}
        \let\@evenfoot\@oddfoot \overfullrule 3pt
        \let\label=\draftlabel
        \let\marginnote=\draftmarginnote
   \def\@eqnnum{(\theequation)\rlap{\kern\marginparsep\tt\@eqnlabel}%
\global\let\@eqnlabel\@vacuum}  }

\makeatletter
\newdimen\normalarrayskip              
\newdimen\minarrayskip                 
\normalarrayskip\baselineskip
\minarrayskip\jot
\newif\ifold             \oldtrue            \def\new{\oldfalse}
\def\arraymode{\ifold\relax\else\displaystyle\fi} 
\def\eqnumphantom{\phantom{(\theequation)}}     
\def\@arrayskip{\ifold\baselineskip\z@\lineskip\z@
     \else
     \baselineskip\minarrayskip\lineskip2\minarrayskip\fi}
\def\@arrayclassz{\ifcase \@lastchclass \@acolampacol \or
\@ampacol \or \or \or \@addamp \or
   \@acolampacol \or \@firstampfalse \@acol \fi
\edef\@preamble{\@preamble
  \ifcase \@chnum
     \hfil$\relax\arraymode\@sharp$\hfil
     \or $\relax\arraymode\@sharp$\hfil
     \or \hfil$\relax\arraymode\@sharp$\fi}}
\def\@array[#1]#2{\setbox\@arstrutbox=\hbox{\vrule
     height\arraystretch \ht\strutbox
     depth\arraystretch \dp\strutbox
     width\z@}\@mkpream{#2}\edef\@preamble{\halign
\noexpand\@halignto
\bgroup \tabskip\z@ \@arstrut \@preamble \tabskip\z@ \cr}%
\let\@startpbox\@@startpbox \let\@endpbox\@@endpbox
  \if #1t\vtop \else \if#1b\vbox \else \vcenter \fi\fi
  \bgroup \let\par\relax
  \let\@sharp##\let\protect\relax
  \@arrayskip\@preamble}
\def\eqnarray{\stepcounter{equation}%
              \let\@currentlabel=\theequation
              \global\@eqnswtrue
              \global\@eqcnt\z@
              \tabskip\@centering
              \let\\=\@eqncr
 \halign to \displaywidth\bgroup
    \eqnumphantom\@eqnsel\hskip\@centering
    $\displaystyle \tabskip\z@ {##}$%
    \global\@eqcnt\@ne \hskip 2\arraycolsep
         $\displaystyle\arraymode{##}$\hfil
    \global\@eqcnt\tw@ \hskip 2\arraycolsep
         $\displaystyle\tabskip\z@{##}$\hfil
         \tabskip\@centering
    &{##}\tabskip\z@\cr}
\begingroup\ifx\undefined\newsymbol \else\def\input#1 \endgroup\fi

\newfont{\hr}{msbm10}
\newfont{\ams}{msam10}
\font\teneufm=cmmib10
\font\seveneufm=cmmib7
\font\fiveeufm=cmmib5
\def\bfit#1{{\textfont1=\teneufm\scriptfont1=\seveneufm
\scriptscriptfont1=\fiveeufm
\mathchoice{\hbox{$\displaystyle#1$}}{\hbox{$\textstyle#1$}}
{\hbox{$\scriptstyle#1$}}{\hbox{$\scriptscriptstyle#1$}}}}

\font\numbers=cmss12
\font\upright=cmu10 scaled\magstep1
\def\stroke{\vrule height8pt width0.4pt depth-0.1pt}
\def\topfleck{\vrule height8pt width0.5pt depth-5.9pt}
\def\botfleck{\vrule height2pt width0.5pt depth0.1pt}
\def\Zmath{\vcenter{\hbox{\numbers\rlap{\rlap{Z}\kern 0.8pt\topfleck}\kern 2.2pt
                   \rlap Z\kern 6pt\botfleck\kern 1pt}}}
\def\Qmath{\vcenter{\hbox{\upright\rlap{\rlap{Q}\kern
                   3.8pt\stroke}\phantom{Q}}}}
\def\Nmath{\vcenter{\hbox{\upright\rlap{I}\kern 1.7pt N}}}
\def\Cmath{\vcenter{\hbox{\upright\rlap{\rlap{C}\kern
                   3.8pt\stroke}\phantom{C}}}}
\def\Rmath{\vcenter{\hbox{\upright\rlap{I}\kern 1.7pt R}}}
\def\Z{\ifmmode\Zmath\else$\Zmath$\fi}
\def\Q{\ifmmode\Qmath\else$\Qmath$\fi}
\def\N{\ifmmode\Nmath\else$\Nmath$\fi}
\def\C{\ifmmode\Cmath\else$\Cmath$\fi}
\def\R{\ifmmode\Rmath\else$\Rmath$\fi}

\def\d{\partial}

\def\bea{\begin{eqnarray}}
\def\eea{\end{eqnarray}}

\def\beq{\begin{equation}}
\def\eeq{\end{equation}}
\def\ba{\beq\new\begin{array}{c}}
\def\ea{\end{array}\eeq}
\def\be{\ba}
\def\ee{\ea}
\def\stackreb#1#2{\mathrel{\mathop{#2}\limits_{#1}}}
\def\Tr{{\rm Tr}}
\def\res{{\rm res}}
\def\Bf#1{\mbox{\boldmath $#1$}}

\def\bPhi{{\Bf\Phi}}

\def\bsigma{{\bfit\sigma}}

\def\2{{1\over 2}}
\def\N2{${\cal N}=2$}
\def\4N{${\cal N}=4$}
\def\1N{${\cal N}=1$}

\def\vpint{{\lefteqn{\int}{\,-}}}
\def\CG{{\cal G}}

\def\half{{\textstyle{1\over2}}}
\def\ha{{1\over 2}}

\def\pint{{-\!\!\!\!\!\!\int}}
\newcommand{\rf}[1]{(\ref{#1})}

\renewcommand{\theequation}{\thesection.\arabic{equation}}

\begin{document}
\begin{flushright}
FIAN/TD-05/04\\
ITEP/TH-20/04
\end{flushright}
\vspace{0.5cm}

\begin{center}
\renewcommand{\thefootnote}{${\!}^\star$}
{\Large \bf Quasiclassical
Geometry and Integrability of AdS/CFT Correspondence
\footnote{
Based on the talks presented at the conferences
{\em Classical and quantum integrable systems},
January 2004, Dubna, and {\em Quarks-2004}, May 2004, Pushkinskie Gory,
Russia} }
\end{center}
\vspace{0.5cm}

\setcounter{footnote}{0}
\begin{center}
{\large A.~Marshakov}\\
\bigskip
{\em Lebedev Physics Institute \&
ITEP, Moscow, Russia}\\
\medskip
{\sf e-mail:\ mars@lpi.ru, mars@itep.ru}\\
\bigskip\bigskip\medskip
\end{center}

\begin{quotation}
\noindent
We discuss the quasiclassical geometry and integrable systems
related to the gauge/string duality. The
analysis of quasiclassical solutions to the Bethe anzatz equations
arising in the context of the AdS/CFT correspondence is performed,
compare to stationary phase equations for the matrix integrals.
We demonstrate how the underlying geometry is related to the
integrable sigma-models of dual string theory, and investigate
some details of this correspondence.
\end{quotation}

\section{Introduction}

Gauge/string duality is an old fascinating subject \cite{Polyakov},
based, in particular, on relationship between the string worldsheets and
Yang-Mills Feynman diagrams at strong coupling. A well-known  example
of such a duality is the matrix model
description of  two-dimensional quantum gravity and low-dimensional
noncritical strings \cite{David:nj,Kazakov:ds}.

A more complicated case of this duality is the so called AdS/CFT correspondence
-- an asserted equivalence of a four-dimensional \4N supersymmetric gauge
theory with a type IIB string theory in the ten-dimensional
$AdS_5\times S^5$ background. Among other predictions, the AdS/CFT conjecture
relates the dimensions of gauge-invariant operators
with the energies of particular closed string states propagating
in the ten-dimensional $AdS_5\times S^5$ spacetime background
\cite{Maldacena:1998re,Gubser:1998bc,Witten:1998qj}.

These anomaluos dimensions in supersymmetric Yang-Mills theory (SYM)
or energies of string states
appear to be the quantities of particular interest, since they can be
sometimes evaluated on different sides of the gauge/string duality,
providing in this sense a quantitative test of the AdS/CFT correspondence.
At present, there is a various number of methods and
approaches for such tests, as well as vast literature on the subject,
but below we are going to concentrate only on a particular way of testing the
equality of the string energies to anomalous dimensions of the
gauge operators.

This way is based on appearence of integrable systems on both sides of the
AdS/CFT correspondence. On gauge side this is {\em quantum} integrable
models, observed even in the context of non-supersymmetric QCD and,
in particular, responsible for diagonalization of the mixing matrix
of renormalization of the constituent operators \cite{Lipatov,Braun}. A
particular simple form \cite{MiZa} this matrix acquires in the sector of
scalar operators, which are absent in non-supersymmetric gauge theories
(generalized set of the famous
BMN operators \cite{BMN}, see e.g. \cite{Gross} on gauge-theory
calculations of their
renormalization). When restricted to the subsector of operators,
consisting of two (among three) complex
holomorphic scalars, it becomes literally equivalent
to the Hamiltonian of the Heisenberg
spin chain, solved long ago via the Bethe anzatz \cite{Bethe,Faddeev}.

On string side this is a classically integrable string
sigma-model, satisfying the world-sheet Virasoro constraints
(unfortunately nothing
is really known yet about the quantum theory on this side, see
however \cite{Schwarz}). Only in particular so-called "pp-wave" limit
of the $AdS_5\times S^5$ background, the
world-sheet Green-Schwarz theory can be quantized in the
light-cone gauge \cite{Metsaev}.
However, this formulation is based on use of the Fock space of
two-dimensional {\em massive} oscillators, which is hardly consistent with
two-dimensional conformal invariance and, is rather a technically
convenient way to describe a collection of effective
{\em classical} oscillator-like modes of an
integrable string model (see \cite{GuKlePo}).

Based on quite a number of achievements in computations of particular
examples of anomalous dimensions \cite{nemcy}
and energies of classical string
trajectoris \cite{tseytlin}, as well as matching between the higher charges
of the entire integrable structures
for the elliptic solutions \cite{AS},
in \cite{KMMZ} a general
approach to comparing the gauge and string
integrable systems was proposed.
Up to now this approach was developed only
for restricted sector of particular scalar operators and classical motions of
string in the (subsector of) compact sphere $S^5$ of the $AdS_5\times S^5$
geometry. It directly relates the {\em quasiclassical}
solutions to Bethe equations, where
elementary magnon excitations form condensates or "Bethe strings", with the
collective classical configurations of spin chains, which can be identified
with the particular limit of the the string sigma-model's finite-gap
solutions.

In these notes we are going to review
the results of \cite{KMMZ} and present
some of them in more general and invariant geometric way.
As a toy model of solving quasiclassicaly
the Bethe anzatz equations we first discuss
the more simple quasiclassical solutions of matrix models. Then we present
the comparison of gauge and string geometry underlying the correspondence.
Finally, we discuss some
physical consequences of the derived formulas and open problems.

\section{Duality and geometry in matrix models}

\begin{figure}[tp]
\centerline{\epsfig{file=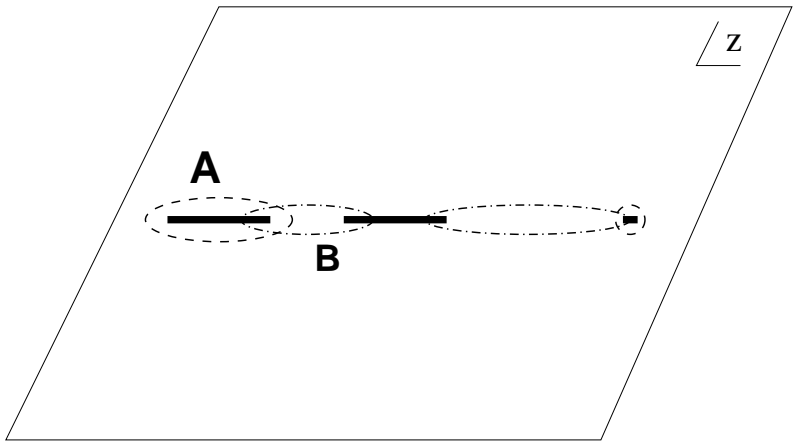,width=105mm,angle=0}}
\caption{Riemann surface of the one-matrix model \rf{mamo}
as a double cover of the $z$-plane. The
eigenvalues are condensed along the cuts, surrounded by $A$-cycles. The
partition funcion (in leading in ${1\over N}$ or quasiclassical
approximation) is defined by its derivatives over the fractions of
eigenvalues
equal to the integrals along the dual $B$-cycles.}
\label{fi:cuts}
\end{figure}

In contrast to higher-dimensional field theories, the zero-dimensional
Yang-Mills theories -- matrix models do not
have renormalizations of any operators and the main aim there is to
compute the (generalized) partition function -- a generating function for
the correlators
\footnote{Here and everywhere below we consider only the
gauge-invariant single-trace operators in matrix theories.}.
The partition function of a matrix model
\be
\label{mamo}
Z = \int d\Phi e^{-{1\over\hbar}{\rm Tr} W(\Phi)} =
{1\over N!}\int \prod_{i=1}^N \left(dz_i
e^{-{1\over\hbar}W(z_i)}\right)
\prod_{i<j}(z_i-z_j)^2
\ee
with some potential $W(\Phi) = \sum_{k=1}^{L+1} t_k\Phi^k$
is related to free energy of dual string theory
\be
\label{expan}
{\cal F} = -\log Z = \sum_{g=0}^\infty \hbar^{2g-2}{\cal F}_{g}
\ee
by the quasiclassical expansion in $\hbar$, which is equivalent to the
't Hooft's ${1\over N}$-expansion, provided by fixing $\hbar N=t_0$.
The quasiclassical solution at $\hbar\to 0$, corresponding
therefore to the planar limit $N\to\infty$ in summing over diagrams,
where all closed string loops or higher
topologies in \rf{expan}
are suppressed, can be found
studying the extrema of effective potential in \rf{mamo}
\be
\label{mamoeq}
W'(z_j) = 2\hbar\sum_{k\neq j}{1\over z_j-z_k}
\ee
If $\hbar=0$ the interaction
is switched off, and all eigenvalues $z_j$
are somehow distributed over the minima where $W'(z)=0$.
Introducing at $N\to\infty$ the eigenvalue density
\be
\label{mamorho}
\rho(z) = \hbar\sum_{j=1}^N\delta(z-z_j), \ \ \
\int_{\bf C} dz\rho(z)=\hbar N\equiv t_0
\ee
or resolvent, defined on the double cover of $z$-plane, cut along
some segments ${\bf C}=\bigcup_j{\bf C}_j$, see fig.~\ref{fi:cuts}
\be
\label{mamoG}
G(z) = \int_{\bf C} {\rho(\zeta)d\zeta\over z-\zeta},\ \ \ \
{1\over 2\pi i}\oint_{\bf C} dzG(z)= t_0
\ee
equation \rf{mamoeq} can be reduced to the integral equation
\be
\label{inteqm}
W'(z) = \vpint_{\bf C}{\rho(\zeta)d\zeta\over z-\zeta} =
G(z_+)+G(z_-), \ \ \ \ z\in\forall{\bf C}_j \subset{\bf C}
\ee
where $z_\pm=z\pm i0$ are two "close" points on two different sides of
the cut -- above and below if cut is along the real axis like it can
happen for polynomial potentials. This equation holds in any point
of the eigenvalue support ${\bf C}$, consisting of several disjoint pieces,
and therefore formula \rf{inteqm} can
be further rewritten as an algebraic equation on the resolvent
$G$ (see e.g. \cite{David})
\be
\label{mamocu}
G^2-W'(z)G=f(z)
\ee
where $f(z)$ in the r.h.s. is a polynomial of the power $L-1$ or
one less than that of $W'(z)$.
Equation \rf{mamocu} defines a hyperelliptic curve and the
quasiclassical free energy ${\cal F}_0$ (the first term in \rf{expan}) can
be entirely defined in terms of the curve \rf{mamocu} and the
generating differential
$Gdz$ \rf{mamoG}. For polynomial potentials the
curve \rf{mamocu} is algebraic curve of finite genus $L-1$ and equation
\rf{mamocu} defines resolvent $G$ as {\em algebraic function} on the
double-cover of $z$-plane. However,
if one allows all possible long operators $\Tr\Phi^L$ for $L\to\infty$
(whose renormalization in four-dimensional theory will be considered below)
it becomes a curve of infinite genus and nothing can be said about the
resolvent $G$ immediately.

Indeed, introducing $Y=W'(z)-2G$ and rewriting \rf{mamocu} as
$Y^2=W'(z)^2+4f(z)$
one immediately finds that this is a genus $g=L-1$ hyperelliptic Riemann
surface
and the auxiliary $L=g+1$ parameters -- the coefficients of the polynomial
$f(z)$ can be
"eaten" by the fractions of eigenvalues on the cuts
\be
S_j = {i\over 4\pi}\oint_{A_j} Ydz = {1\over 2\pi i}\oint_{A_j} Gdz =
\int_{{\bf C}_j}\rho(z)dz,
\ \ \ \ \ j=1,\dots,L-1
\ee
together with their total number $t_0$ (see \rf{mamoG}), and the dependence
of free energy upon these variables is given by
$${\d {\cal F}_0\over\d S_j}=
\ha\oint_{B_j} Ydz
$$
where dual contours $B_j$ are drawn on fig.~\ref{fi:cuts}.

Finally in this section let us point out that for non-polynomial potentials,
when $L\to\infty$, to get any reasonable answer
one should consider the situation when only finite number of extrema of
the potential $W'(x)=0$ are filled in by eigenvalues. The number of
condensates $K$ then becomes an extra parameter of the problem and, generally
speaking, one should
consider any $K<L$. This is in quite direct analogy with what are going to
do below: for the long operators (nontrivial analogs of $\Tr \Phi^L$ for
$L\to\infty$) one has to define the Riemann surface "by hands", and not
simply using the quasiclassical Baxter curve of, as in
\cite{KorchQ}, -- an analog of the
matrix model curve \rf{mamocu} for a non-polynomial potential.

\section{SYM and geometry of quasiclassical Bethe equations}

In contrast to matrix model the AdS/CFT conjecture deals
with the {\em four-dimensional} \4N SYM with the $SU(N)$ gauge group.
Again, the main contribution in ${1\over N}$-expansion, where
closed string loops are suppressed, comes
from the planar diagrams when $N\to\infty$  at fixed 't Hooft coupling
$\lambda=g_{YM}^2N=g_sN$, analogous to $t_0$ of the previous section,
while string coupling $g_s=g_{YM}^2$ is an analog of
quasiclassical parameter $\hbar$.
At $\lambda \gg 1$ the \4N SYM theory is beleived to be
dual to string theory in $AdS_5\times S^5$ with the equal radii of
curvature ${R\over\sqrt{\alpha'}}=\lambda^{1/4}$. Therefore any
test of the AdS/CFT conjecture
implies comparing analytic series at $\lambda=0$
(SYM perturbation theory) with analytic in
$\alpha'\propto{1\over\sqrt{\lambda}}$
worldsheet expansion
\footnote{There were, however, some attempts (not very promissing from our
point of view) to consider the world-sheet
theory around so called "null-string" limit with
$\alpha'\to\infty$, see \cite{Mikhailov}.}.

A possible way-out from this discrepancy in parameters of
expansion can be to consider the
{\em classical} string solutions with large values of integrals of motion
(usually referred as "spins" $J$) on $AdS_5\times S^5$ side
\cite{GuKlePo,tseytlin}, whose energies
should correspond to anomalous dimensions of
"long" operators on gauge side. In this case the classical string energy
of the form $\Delta = \sqrt{\lambda}{\cal E}
\left({J\over\sqrt{\lambda}}\right)$ may have an expansion of the form
$\Delta = J + \sum_{l=1}^\infty E_l\left({\lambda\over J^2}\right)^l$ over
the integer powers of 't Hooft coupling, which can be treated as series
at $\lambda=0$ even at $\lambda\gg 1$ provided large $\lambda$ is suppressed
by large value of the integrals of motion $J$.
If it happens (this is not, of course, guaranteed) the classical
string energy can be tested by direct comparison with perturbative
series for gauge theory.

The four-dimensional \4N SYM is conformal theory, i.e. $\beta (g_{YM})=0$,
but the anomalous dimensions $\gamma$ of the
composite operators, e.g. $\Tr \left(\Phi_{i_1}\dots\Phi_{i_L}\right)$
are still renormalized nontrivially. In \cite{KMMZ} and below we consider
the particular scalar operators from this set, though the proposed
approach can be applied in much more general situation. On string side
such operators correspond to the string motion in the compact $S^5$-part
of ten-dimensional target-space, due to standard Kaluza-Klein argument.
To simplify the situation maximally,
choose two complex $\bPhi_1=\Phi_1+i\Phi_2$ and $\bPhi_2=\Phi_3+i\Phi_4$
fields
among six real $\Phi_i$ and consider the {\em holomorphic} operators
\be
\label{holop}
{\rm Tr}\left( \bPhi_1 \bPhi_1 \bPhi_1 \bPhi_2 \bPhi_2 \bPhi_1 \bPhi_2 \bPhi_2
\bPhi_1 \bPhi_1 \bPhi_1 \bPhi_2 \ldots\right)
\ee
which can be conveniently labeled by arrows as
$
\left|\uparrow\uparrow\uparrow\downarrow\downarrow\uparrow
\downarrow\downarrow\uparrow\uparrow\uparrow\downarrow\ldots
\right\rangle \in \left({\bf C}^2\right)^{\otimes L}
$
The holomorphic subsector is "closed" under renormalization and
anomalous dimensions are eigenvalues of the $2^L\times 2^L$ mixing
matrix
\be
\label{hahe}
H=\frac{\lambda}{16\pi^2}\sum_{l=1}^L
\left(1-{\sigma}_l\cdot{\sigma}_{l+1}\right)+O(\lambda^2)
\ee
which is, up to addition of a constant, the permutation operator in
$\left({\bf C}^2\right)\otimes
\left({\bf C}^2\right)$, whose appearence is determined by structure of the
$\Phi^4$-vertex in SYM Lagrangian,
or the Hamiltonian for Heisenberg magnetic \cite{MiZa}.

It is well-known, that this matrix
can be diagonalized using the Bethe anzatz \cite{Bethe}, (see e.g.
\cite{Faddeev} for present status of this technique and
comprehensive list of references).
Eigenvectors of $H$
are parameterized by Bethe roots $\{ u_1,\dots, u_J\}$, ($J \leq \half L$
due to an obvious $\Z_2$-symmetry) satisfying
the Bethe ansatz equations for the Heisenberg spin chain
\be\label{BAE}
L\log\left(u_j+i/2\over u_j-i/2\right)=2\pi in_j
+\sum_{k\ne j}^J \log{u_j-u_k+i\over u_j-u_k-i}
\ee
(with mode numbers $n_j\in\Z$ being some integers, never vanishing
for a nontrivial solution). For the operators \rf{holop} equations \rf{BAE}
are supplied by the "trace condition"
\be
\label{MOMu}
e^{iP}=\prod_{j=1}^J{u_j+i/2\over u_j-i/2}=1,
\ee
i.e. integrality of the total momentum: ${P\over 2\pi}\in
{\bf Z}$.  The energy of the solution, is given by sum over magnons, i.e. in
the leading order
\be
\label{gamad}
\gamma={\lambda\over 8\pi^2}\sum_{j=1}^J{1\over u_j^2+1/4} + O(\lambda^2)
\ee
and equals to the (one-loop)
Yang-Mills scaling dimension $\gamma$ up to a factor.

\begin{figure}[tp]
\centerline{\epsfig{file=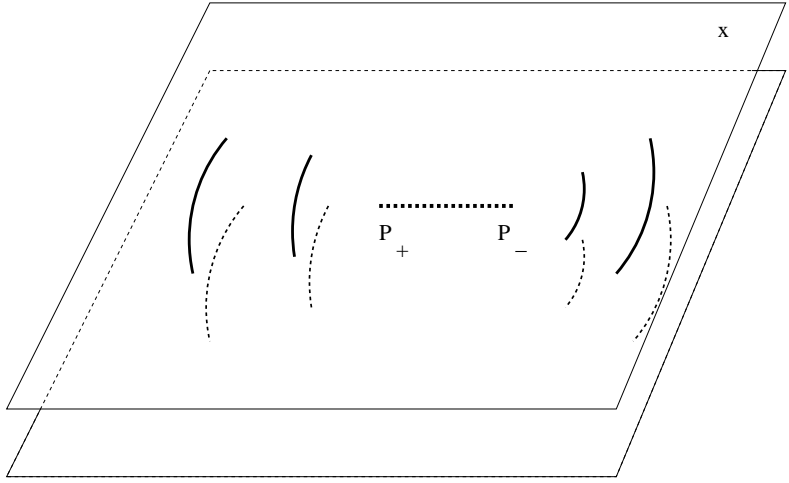,width=75mm,angle=-90}}
\caption{Riemann surface $\Sigma$, which is a double-cover of the $x$-plane
cut along the Bethe strings (four slightly curved lines on each sheet in this
example), which cross the real axis at $x={1\over 2\pi n_l}$.
The upper sheet is physical, while on the lower sheet the
resolvent may have extra singularities.}
\label{fi:sigma}
\end{figure}

For comparison with dual string theory we are interested in the
long operators with $L\to\infty$, for which the Bethe roots are
typically of the order of $u_j\sim L$. Rescaling $u_j = Lx_j$, and
neglecting the higher in ${1\over L}$ terms, one gets from \rf{BAE}
\be\label{BAEG}
{1\over x_j}=2\pi n_j+ {2\over L}\sum_{k\ne j}^J {1\over
x_j-x_k}
\ee
Like in the matrix model case \rf{mamoeq}, in the absence of the
second in the r.h.s. of
\rf{BAEG} "interaction term", $x_j= {1\over 2\pi n_j}$ for each $n_j$, and
when we switch on the interaction
the roots corresponding to $n_j$ will "concentrate" around
${1\over 2\pi n_j}$ and form the so called "Bethe strings", shown at
fig.~\ref{fi:sigma}.
Introduce at $L\to\infty$, as in \rf{mamorho}, \rf{mamoG}, the density
\be
\label{rhoJL}
\rho(x)=\frac{1}{L}\sum_{j=1}^J\delta(x-x_j), \ \ \ \
\int_{\bf C} dx\rho(x)={J\over L}
\ee
or resolvent
\be
\label{defG}
G(x ) = {1\over L}\sum_{j=1}^J {1\over x-x_j} =
\int_{\bf C}{d\xi\,\rho(\xi)\over x-\xi}, \ \ \ \
{1\over 2\pi i}\oint_{\bf C} dxG(x)={J\over L}
\ee
Suppose we have {\em finite} number of different $n_l\neq n_{l'}$, with
$l,l'=1,\dots,K$, then in the scaling limit
the total eigenvalue support is again
${\bf C}={\bf C}_1\bigcup\ldots\bigcup{\bf C}_K$, where on each component
one gets from \rf{BAEG}
\be\label{BAEC}
2\pint_{{\bf C}} {d\xi\,\rho(\xi)\over x-\xi} =
G(x_+)+G(x_-) = {1\over x}-2\pi n_l,\ \ \ x\in
{\bf C}_l
\ee
where $G(x_\pm)$ are values of the resolvent on two different sides of the
cut.
The integrality of total momentum condition, using the Bethe
equations \rf{BAEG}, acquires the form
\be
\label{MOMx}
{1\over L}\sum_{j=1}^J {1\over x_j} =
2\pi \sum_{l=1}^K n_l \int_{{\bf C}_l}\rho(x)dx =
2\pi m, \ \ \ \ m\in \Z
\ee
or
\be
\label{MOMrho}
{1\over 2\pi i}\oint_{{\bf C}}{G(x)dx\over x}
= 2\pi m, \ \ \ \ \ n_l, m\in \Z.
\ee
Different $n_l\neq n_{l'}$ on different parts of support ${\bf C}_l\bigcap{\bf
C}_{l'}=\emptyset$ mean that, in contrast to the matrix model case
$G(x)=\int^x dG$ is not already a single-valued function, but an
Abelian integral on some hyperelliptic curve $\Sigma$
\be
\label{sigmaxxx}
 y^2 = R_{2K}(x) = x^{2K} +
r_1x^{2K-1} + \dots + r_{2K} =
\prod_{j=1}^{2K} (x-{\rm x}_j)
\ee
where ${\rm x}_j$ are roots of the polynomial $R_{2K}(x)$.
Equations \rf{defG}, \rf{BAEC} and \rf{MOMrho} can be solved after
reformulating them as a set of properties of the meromorphic
differential $dG$:
\begin{itemize}
  \item $dG$ is the second-kind Abelian differential with the only
second-order pole at the point $P_0$, ($x(P_0)=0$ on unphysical
sheet of the Riemann surface $\Sigma$);
  \item $dG$ has integral $B$-periods
\be
\label{Bint}
\oint_{B_i}dG = 2\pi (n_i-n_K)
\ee
More exactly one can write \cite{KMMZ}
\be
\label{B'int}
\int_{B'_j}dG = 2\pi n_j, \ \ \ \ j=1,\dots,K+1
\ee
where $B'_j$ is the contour from $\infty_-$ on the lower sheet to
$\infty_+$ on the upper sheet, passing through the $j$-th cut, so that
$B_j=B'_j-B'_K$, for $j=1,\dots,K$;
  \item $dG$ has the following behaivior at infinity
\be
\label{inf}
dG\ \stackreb{x\to\infty}{=}\ {J\over
L}{dx\over x^2}+\dots
\ee
and the Abelian integral $G(x)$ itself is fixed by
\be
\label{zero}
G(x) = 2\pi m + \int_0^x dG, \ \ \  {\rm or}\ \  G(0)=2\pi m
\ee
\end{itemize}
The general solution for the differential $dG$ on hyperelliptic curve
\rf{sigmaxxx}, satisfying the above requirements
\rf{Bint}, \rf{B'int}, \rf{inf} and \rf{zero} reads \cite{KMMZ}
\be
\label{gsol}
dG =
-{dx\over 2x^2}\left(1-{\sqrt{r_{2K}}\over y}\right)
+ {r_{2K-1}\over 4\sqrt{r_{2K}}}{dx\over xy}
+ \sum_{k=1}^{K-1}a_k {x^{k-1}dx\over y}
\ee
together with the extra conditions, ensuring single-valuedness of the
resolvent on "upper" physical sheet
\be
\label{Aint}
\oint_{A_i}dG = 0,\ \ \ \   i=1,2,\ldots,K-1
\ee
to be easily solved for the coefficients $\{ a_k \}$.
The rest of parameters is "eaten by" fractions of roots on particular
pieces of support
\be
\label{frac}
S_j = -{1\over 2\pi i}\oint_{A_j}x dG =
\int_{{\bf C}_j} \rho(x)dx, \\ j=1,\dots,g=K-1
\ee
the total amount of Bethe roots \rf{rhoJL},
and the total momentum \rf{MOMrho}.

The energy or one-loop anomalous dimension for generic finite-gap
solution \cite{KMMZ} can be read from \rf{gamad}, \rf{gsol}
\be
\label{gamma}
\gamma = {\lambda\over 8\pi^2 L}\oint_{\bf C} {dx\over 2\pi i\, x^2}\, G(x)=
{ \lambda\over 8\pi^2L  }\left({r_{2K-2}\over 4 r_{2K}}
-\frac{r^2_{2K-1}}{16r^2_{2K}}
-{a_1\over \sqrt{r_{2K}}}\right)
\ee
The anomalous dimensions defined by \rf{gamma} are functions
of the coefficiets of the embedding equation \rf{sigmaxxx} and
$a_1$ which again
is expressed through these coefficients by means of \rf{Aint}. The
moduli of the curve \rf{sigmaxxx} are themselves
(implicitly) expressed through the mode numbers $n_j$ and
root fractions
$S_j$ via \rf{Bint} or \rf{B'int} and \rf{frac} (together
with the total momentum
\rf{MOMrho} and the total number of Bethe roots \rf{defG}).

\section{Geometry of classical string solutions}

The general solution for anomalous dimension \rf{gamma} is
expressed through the
integrals of motion on some {\em classical} configurations of the Heisenberg
magnet \cite{KMMZ}
\footnote{Such correspondence with classical solutions
was first noticed in \cite{RESHSM}
for the non-linear Schr\"odinger equation.}.
In the dual string picture one has the classical trajectories of
string,
moving in (subspace of) $AdS_5\times S^5$ and the finite gap
solutions to string sigma-model in AdS-like spaces were first
constructed in
\cite{Krisig}. In the Appendix to \cite{KMMZ} it was demonstrated that
this construction, slightly modified, can be easily
applied to the case of compact $S^d$ sigma-models. In this
section we are going to show how these classical solutions can be compared with
the quasiclassical solutions on the gauge side.

In particular subsector of only two holomorphic fields one gets the
$S^3\subset S^5$ sigma-model (in the $AdS_5$-sector the only nontrivial
string co-ordinate on the solution is "time"
$X_0={\Delta\over\sqrt{\lambda}}\ \tau$),
which is equivalent \cite{FR} (since $S^3$
is the group-manifold of $SU(2)$)
to the $SU(2)$ principal chiral field  with the Lax pair,
(see \cite{KMMZ} for more details)
\be
J_\pm(x) =
{\Delta\over\sqrt{\lambda}}\frac{i{\bf S}_\pm\cdot{\bsigma}}{1\mp X}
\\
\d_+J_--\d_-J_++[J_+,J_-]=0
\\
\d_+J_-+\d_-J_+=0
\label{zerocur}
\ee
which has two simple poles at values of string spectral
parameter $X= X(P_\pm)=\pm {\sqrt{\lambda}\over 4\pi\Delta}$. In different
words, such sigma-model is equivalent
to a system of {\it two} interacting relativistic spins
${\bf S}_+$ and ${\bf S}_{-}$:
\be
\label{emcf}
\d_+{\bf S}_{-}+{2\Delta\over\sqrt{\lambda}}\ {\bf S}_{-}\times{\bf S}_{+}=0,
\\
\d_-{\bf S}_{+}-{2\Delta\over\sqrt{\lambda}}\ {\bf S}_{-}\times{\bf S}_{+}=0.
\ee
which
in some "non-relativistic limit" degenerates into the Heisenberg magnet
\cite{KMMZ}, which is similar to a limit,
studied in the papers \cite{Kruczenski}.

\begin{figure}[tp]
\centerline{\epsfig{file=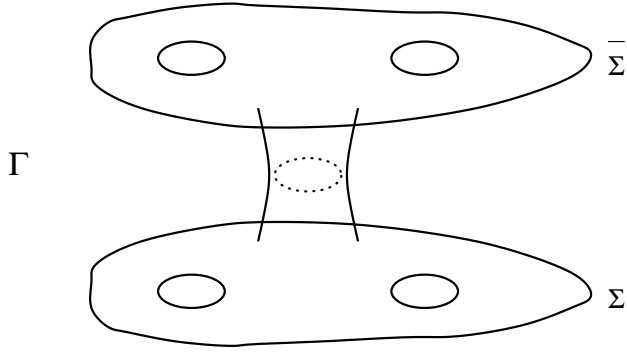,width=75mm,angle=-90}}
\caption{Riemann surface $\Gamma$, which is a double cover of $\Sigma$
with a single cut.}
\label{fi:gamma}
\end{figure}

However, this method can be used only for the group-manifolds. Nevertheless,
in general situation the string sigma-model solution for the complex
co-ordinates
$Z_I(\tau,\sigma)$ and $\bar Z_I(\tau,\sigma)$ on $S^{2D-1}$, (constraint
by $\sum_I |Z_I|^2=1$)
\be
\label{zkri}
Z_I(\sigma_\pm) = r_I\Upsilon(q_I,\sigma_\pm)
\ \ \ \ \
{\bar Z}_I(\sigma_\pm) = r_I\Upsilon({\bar q}_I,\sigma_\pm)
= r_I\overline{\Upsilon(q_I,\sigma_\pm)},
\ \ \ \ \
I=1,\dots,D
\ee
can be found \cite{Krisig}
in terms of the Baker-Akhiezer (BA) functions
\be
\label{BAsigma}
\Upsilon(P,\sigma_\pm)\
\stackreb{P\to P_\pm}{=}\ e^{k_\pm \sigma_\pm}
\left(1+\sum_{j=1}^\infty {\xi_j(\sigma_\pm)\over k_\pm^j}\right)
\propto e^{\Omega_+(P)\sigma_+
+ \Omega_-(P)\sigma_-}
\theta \left({\bf{\cal A}}(P)+ {\bf U}_+\sigma_+
+ {\bf U}_-\sigma_-\right)
\ee
defined on double cover $\Gamma$ (branched at $P_+$ and $P_-$)
of a Riemann surface $\Sigma$ (see fig.~\ref{fi:gamma}).
For only two complex co-ordinates $Z_I$ (like in the $S^3$
case) the curve $\Sigma$ is hyperelliptic and directly related with the
curve \rf{sigmaxxx} of the Heisenberg chain. For the $S^5$ case
$\Sigma$ can be presented as some three-sheet cover of an $X$-plane,
and is presumably related to the three-sheet covers arising in solving of
the Bethe anzatz equations for the operators beyond the $SU(2)$ sector
\cite{Kristj,Engquist,Minahan}.

The BA function \rf{BAsigma} (and hence the solution to sigma-model)
is constructed in terms of {\em two}
second-kind Abelian differentials $d\Omega_\pm$ on Riemann surface $\Gamma$
\be
d\Omega_\pm\ \stackreb{P\to P_\pm}{=}\
\pm dk_\pm\left(1+O(k_\pm^{-2})\right),
\ \ \ \ \ \oint_{\bf A}d\Omega_\pm = 0
\ee
with the only second-order pole at
$P_\pm$ respectively;
${\bf U}_\pm = \oint_{\bf B} d\Omega_\pm$ are the vectors of
their ${\bf B}$-periods.

The proof of the fact that formulas \rf{zkri} are solutions to the sigma-model,
satisfying classical Virasoro constraints, is based on existence of
the third-kind Abelian differential $d\Omega$ on $\Sigma$
with the simple poles at $P_\pm$ and zeroes in the poles
of the BA functions $\Upsilon$ and conjugated $\bar\Upsilon$. Then
one may define the string resolvent or quasimomentum by the following
formula
\be
\label{gstr}
d\CG = \ha\left(d\Omega_+ - d\Omega_-\right)
\\
d\CG\ \stackreb{P\to P_\pm}{=}\
\ha dk_\pm\left(1+O(k_\pm^{-2})\right),
\ \ \ \ \ \oint_{\bf A}d\CG = 0
\ee
For periodic in $\sigma$ solution, as follows from \rf{BAsigma},
the $B$-periods of the resolvent
${1\over 2\pi}\oint_{\bf B} d\CG \in\Z$ are integer-valued,
and for the periodic solutions one can write
\be
\label{domega}
d\Omega = {\bar\Upsilon\Upsilon \over
\langle\bar\Upsilon\Upsilon\rangle}d\CG
\ee
where brackets mean the average over the period in $\sigma$-variable.

The BA function \rf{BAsigma} satisfies the second-order
differential equation
\be
\label{wave}
(\d_+\d_- + u)\Upsilon (P,\sigma_\pm)=0, \ \ \ \ P\in\Gamma
\ee
where $u \propto
\sum_i\left(\d_+ Z_i\d_-\bar Z_i + \d_- Z_i\d_+\bar Z_i\right)$.
This fact and the Virasoro constraints
$\lambda\sum_I\left|\d_\pm Z_I\right|^2 = \Delta$ are guaranteed
by the properties of the differential \rf{domega} and existence
of the function $E$ on $\Sigma$ with $D$ simple poles $\{ q_I\}$ and the
following behaivior at the vicinities of the points
$P_\pm$: $E=E_\pm \pm {4\pi\Delta\over\sqrt{\lambda}}
{1\over k_\pm^2} + \dots$, (cf. with \cite{Krisig}).

The normalization factors in the expressions for the sigma-model
co-ordinates (\ref{zkri}) are determined by the formulas
\be
\label{ris}
r_I^2 = {\res_{q_I} Ed\Omega\over E_--E_+}, \ \ \ \ I=1,\dots,D
\ee
where $E_\pm = E(P_\pm)$ and
normalizations (\ref{ris}) satisfy $\sum_{I=1}^D r_I^2=1$ due to
vanishing of the total sum over the residues $\sum \res
\left(Ed\Omega\right) = 0$.

Rescaling
$\d_\pm\to\frac{\sqrt{\lambda}}{4\Delta}\,\d_\tau\pm \d_\sigma$,
$\Upsilon\to e^{{2\Delta\over\sqrt{\lambda}}\tau}\Psi$,
$\bar\Upsilon\to e^{-{2\Delta\over\sqrt{\lambda}}\tau}\bar\Psi$,
$u\to
u-{4\Delta^2\over\lambda}$,
in the limit $\lambda/\Delta^2 \ll 1$, one gets from \rf{wave} the
non-stationary Schr\"odinger equation
\be
\label{KP}
\left(\d_\tau-\d_\sigma^2 + u\right)\Psi = 0
\\
\left(-\d_\tau-\d_\sigma^2 + u\right)\bar\Psi = 0
\ee
where now both BA functions $\Psi$ and $\bar\Psi$ can be defined on Riemann
surface $\Sigma$ (see fig.~\ref{fi:sigma}), with the ends of the extra cut
$P_\pm$ are shrinked to a single point $P_0$, with the expansion at the vicinity
of this point (with new local parameter $k(P_0)=\infty$)
\be
\label{PsiKP}
\Psi \stackreb{P\to P_0}{=} e^{k\sigma+k^2\tau}\left(1+{\psi_1\over k}+
{\psi_2\over k^2}+
\dots\right)
\\
\bar\Psi \stackreb{P\to P_0}{=} e^{-k\sigma-k^2\tau}\left(1+
{\bar\psi_1\over k}+
{\bar\psi_2\over k^2}+\dots\right)
\ee
Substituting expansions \rf{PsiKP} into \rf{KP} one gets
\be
\label{relKP}
u = 2{\d\psi_1\over\d\sigma}=-2{\d\bar\psi_1\over\d\sigma}
\\
{\d\psi_1\over\d\tau}-2{\d\psi_2\over\d\sigma}-{\d^2\psi_1\over\d\sigma^2}+
u\psi_1=0
\\
-{\d\bar\psi_1\over\d\tau}+2{\d\bar\psi_2\over\d\sigma}-
{\d^2\bar\psi_1\over\d\sigma^2}+
u\bar\psi_1=0
\ee
For the conjugated functions one has $\psi_1+\bar\psi_1=0$ and then
${\d\over\d\sigma}\left(\psi_2+\bar\psi_2+\psi_1\bar\psi_1\right)
=-{\d u\over\d\sigma}$. Therefore, one gets an expansion
\be
\label{pp}
\bar\Psi\Psi \stackreb{P\to P_0}{=} 1 + {u\over k^2} + \dots
\ee
The differential \rf{domega} in this limit turns into
$d\Omega={\bar\Psi\Psi\over \langle\bar\Psi\Psi\rangle} dG$ and function
$E$ acquires a simple pole at $P_0$, i.e. $E \stackreb{P\to P_0}{=}
k+\dots$, if written in terms of new local parameter $k$.
From vanishing of the sum of the residues of differential $Ed\Omega$
one gets now
\be
u = \res_{P_0} FdQ =
-\sum_{q_I} F{\bar\Psi\Psi\over \langle\bar\Psi\Psi\rangle} dG \propto
\sum_I \bar\Psi(q_I)\Psi(q_I)
\ee
It means, that $\Psi_I(\tau,\sigma)\propto\Psi (q_I)$ satisfy some vector
non-linear Schr\"odinger equation \cite{Cherednik}
\be
\label{vschro}
\left(\d_\tau-\d_\sigma^2 + \sum_J|\Psi_J|^2\right)\Psi_I = 0
\ee
In the case of $D=2$ the curve $\Sigma$ is hyperelliptic and one can take
the function $E$ with the only two poles, see below.
Then \rf{vschro} turns into the
ordinary non-linear Schr\"odinger equation, which can be transformed
to the Heisenberg magnetic chain \cite{ZT,FT}:
\be
\label{nlshe}
|\Psi|^2 \propto {\bf S}_\sigma^2
\\
\bar\Psi\d\Psi - \Psi\d\bar\Psi \propto \left( {\bf S}_\sigma\cdot
{\bf S}\times{\bf S}_{\sigma\sigma}\right)
\ee
and so on, which is a gauge transformation for the Lax operators.

Another way to describe classical string geometry was proposed in
\cite{KMMZ} and was based on reformulating of geometric data of the
principal chiral field \rf{zerocur} in terms of some Riemann-Hilbert problem.
The spectral problem on string side (a direct analog of the formulas
\rf{defG}, \rf{BAEC}, \rf{MOMrho} and \rf{gamma})
can be formulated in the following way.
Let $X$ and $\CG(X)$ be string spectral parameter and resolvent,
equal to the quasimomentum of the classical solution
(maybe up to an exact one-form). The spectral Riemann-Hilbert
problem on string side
can be written as \cite{KMMZ}
\be
\label{streq}
{1\over 2\pi i}\oint_{\bf C} \CG(X)dX= {J\over \Delta} + {\Delta-L\over 2\Delta}
\\
{1\over 2\pi i}\oint_{\bf C} {dX \CG(X)\over X} = 2\pi m
\\
\oint_{\bf C} {2tdX\over X^2}{ \CG(X)\over 2\pi i} =  \Delta - L
\ee
and
\be
\label{strbete}
\CG(X_+)+\CG(X_-)-2\pi n_l = {X\over X^2-t}
\ee
where we introduced the notation $t={\lambda\over 16\pi^2\Delta^2}$.

Consider now $x=X+{t\over X}$ as {\em exact} change of spectral parameter,
together with $G(x)={\cal G}(X)$
\footnote{In \cite{KMMZ} it has been introduced in the leading order in
$\lambda$. A similar change of variables has been proposed in \cite{BeSt?},
with $\Delta$ replaced with its "bare value" $L$.}. This is literally an exact
change of the local co-ordinate in the vicinity of an extra cut
in the general construction discussed
above. Indeed, in terms of $k_\pm$ one has $X \stackreb{P\to P_\pm}{=}
\pm\sqrt{t} + {1\over k_\pm}+\dots$, then $x \stackreb{P\to P_\pm}{=} \pm
2\sqrt{t} \pm {1\over\sqrt{t}}{1\over k_\pm^2} + \dots$, i.e. the function
$E = {1\over x}$ (up to an overall constant) satisfies all desired
properties, e.g. when the cut between $P_+$ and $P_-$ on fig.~\ref{fi:sigma}
shrinks to a point $P_0$ with $x(P_0)=0$ the function $E$ acquires a simple
pole at this point.

Proceeding further
\be
dx=dX\left(1-{t\over X^2}\right) = {dX\over X}\left(X-{t\over X}\right)
= {dX\over X}\sqrt{x^2-4t}
\ee
and combination of the first and
third lines in \rf{streq} gives
\be
{1\over 2\pi i}\oint \CG(X)dX= {J\over \Delta} + {\Delta-L\over 2\Delta} =
{J\over \Delta} + t\oint {\CG(X)dX\over 2\pi iX^2}
\ee
or
\be
\label{nogabe}
{1\over 2\pi i}\oint dx G(x)={J\over \Delta}
\ee
The second line of \rf{streq} is then
\be
\label{momgabe}
{1\over 2\pi i}\oint {dx G(x)\over \sqrt{x^2-4t}} = 2\pi m
\ee
where the integral is taken around the cut between the points
$-2\sqrt{t}$ and $2\sqrt{t}$ in the $x$-plane,
and the third line of \rf{streq} gives
\be
\label{dgabe}
\oint {dx G(x)\over 2\pi i}
\left({x\over\sqrt{x^2-4t}}-1\right) = \Delta - L
\ee
The "string Bethe" equation on the cuts \rf{strbete}
turns now into
\be
\label{strgabe}
G(x_+)+G(x_-)-2\pi n_l =
{1\over\sqrt{x^2-4t}}
\ee
We now see from \rf{nogabe}, \rf{momgabe}, \rf{dgabe} and \rf{strgabe} that
the classical string theory spectral problem looks identically to the
quasiclassical Bethe equations on gauge side upon replacements
\be
\label{repl}
{1\over x}\ \rightarrow {1\over\sqrt{x^2-4t}} = {1\over x} + {2t\over x^3}
+\dots
\\
L \rightarrow \Delta
\\
\gamma \rightarrow \Delta - L
\ee
In other words, this leads to a nonlinear relation
\be
\label{strdg}
\Delta - L = \Gamma (\lambda,\Delta)
\ee
where $\Gamma (\lambda,L) = \gamma + O(\lambda^2)$ should be compared with
the multi-loop anomalous dimension of the supersymmetric gauge theory.

A simplest non-trivial example
of such relation is the solitonic limit of small number of Bethe roots,
leading to
the "modified" BMN formula \cite{KMMZ}
\be
\label{BMN}
\Delta-L=\sum_k N_k\left(
\sqrt{1+\frac{\lambda n^2_k}{\Delta^2}}-1\right)
\ee
for $J=\sum_k N_k$ expressed as a total amount of "positive" $n_k>0$ and
"negative" $n_k<0$ massive oscillators \cite{Metsaev}. Formulas \rf{strdg} and
\rf{BMN} show, that the solution for $\Delta$ of classical string theory is
given in terms of the highly non-linear formulas, and the oscillator
language of \cite{Metsaev,BMN} is rather an effective tool for descriprtion
of certain quasiclassical modes of an integrable string model in pp-wave
geometry, than an exact world-sheet quantization.

\section{Discussion}

In these notes we have discussed the recent attempts of quantitative
verification of the AdS/CFT correspondence based on appearence of integrable
structures on both sides of the gauge/string duality. It turns out that the
quasiclassical solution to the Bethe anzatz equations arising in the process
of diagonalization of the mixing matrix for constituent operators can be
formulated in terms of (discrete) families of complex
curves endowed with a generating
one-form, quite similar to the quasiclassical solutions of matrix models
and Seiberg-Witten gauge theories.

In contrast to the matrix model case, the Bethe roots for the "compact"
chains
form strings never lying along the real axis. Moreover, the resolvent for
the infinite number of Bethe roots
cannot be expressed through an algebraic function on a curve
of finite genus due to nontrivial mode numbers $n_j$, whose total
number is fixed
to be finite for the class of finite-gap or algebro-geometric solutions.

The condensation of Bethe roots on the cuts of Riemann surfaces leads to
the fact that for long $L\to\infty$ operators the
corresponding anomalous dimensions are expressed through the integrals of motion
of some {\em classical} spin waves of the corresponding magnetic.
This is a very
nontrivial "continuum limit", since many quantum spins condense into
collective classical mode;
this nontriviality was also discussed in \cite{Kruczenski}, though
the approach presented above is much more simple. Hence,
considering long operators and solving Bethe equations for them we finally
come to some finite-gap solutions determined by particular complex curves.

On the other side of diality one has classically integrable string sigma-model.
Despite there are no arguments why this approximation on string side maybe
valid for comparison with the gauge theory (and this is not true, e.g. for
the operators corresponding to motion of string in the non-compact part
of $AdS_5\times S^5$ \cite{GuKlePo}, see also discussion of this issue
e.g. in \cite{Gorsky}), for
certain solutions nevertheless the large $\lambda$ is
suppressed by large values of the integrals of motion and field-theoretic
perturbation theory can be reproduced from the string calculations.

Generally the string sigma-model (even restricted to bosonic part of
compact $S^5$ or its subspace) is a system with infinitely many degrees of
freedom. However, following \cite{Krisig} it is possible to construct its
finite-gap solutions, satisfying the world-sheet Virasoro constraints.
We have investigated the properties of such solutions and demonstrated that
underlying geometry can be naturally sewed with the quasiclassical geometry
of the Bethe anzatz solutions.

In more exact terms the generic
sigma-model solution is formulated using the complex curve
$\Gamma$, being a simple "one-cut" double cover of some curve $\Sigma$,
where exists a function $E$ with $D$ simple poles. It means that $\Sigma$ can be
considered as a $D$-sheet cover of Riemann sphere, being for $D=2$ (the
$S^3$ case) a hyperelliptic curve. The solution is constructed using the
standard technique of the finite-gap potentials for
non-stationary one-dimensional and two-dimensional
the Schr\"odinger operators. For $D>2$ the appearence of vector non-linear
Sch\"rodinger provides an intuition for what kind of solutions to the nested
Bethe anzatz should be looked for beyond the $SU(2)$ subsector.

However, the "weak-coupling" limit to the gauge side is rather nontrivial.
The geometry of $\Gamma$ changes so that it becomes just two copies of
$\Sigma$ with the extra cut on fig.~\ref{fi:sigma} shrunk to a single point
$P_0$. The sigma-model solution turns into solution of $(D-1)$-dimensional
vector non-linear Sch\"rodinger equation, equivalent for $D=2$ to the
Heisenberg magnetic chain.

The subtleties of this limit were widely discussed in the literature. In
\cite{SS} it was proposed to describe higher loops on the SYM side in terms
of the integrable chain \cite{Ino}, whose transfer matrix is defined on
elliptic curve with one of the periods proportional to the length of the
chain. Such reformulation of perturbative SYM leads to difference with the
predictions of classical string theory
starting from three loops, the most comprehensive discussion of
these discrepancies can be found in \cite{BeSt?}.

It would be interesting to understand how the integrable models, associated
to the elliptic curve, can appear in the framework of
approach discussed in these notes. The
periodicity in the length of the chain $L$ naturally arises when one
computes more than $L$ loops on the gauge side, but this is certainly
inconsistent with taking the $L\to\infty$ limit first. It means that one has
to study the finite-size ${1\over L}$-corrections to the quasiclassical
Bethe equations, see e.g. \cite{Zalast}. From the point of view of generic
sigma-model solutions the periodicity in $\sigma$ comes automatically, when
one considers the curve $\Sigma$ as cover of some elliptic curve, instead of
the $x$-plane or Riemann sphere. This
suggests a possible way of development of the considered problems along the
lines of \cite{KriZa}.

\bigskip\noindent
{\bf Acknowledgements} I am grateful to V.Kazakov, J.Minahan and K.Zarembo
for collaboration in \cite{KMMZ} and to A.Gorsky, A.Mikhailov and, especially,
I.Krichever for important discussions. The work was partially supported by
the RFBR grant 04-01-00642, the INTAS grant 00-561,
the program of support of scientific schools 1578.2003.2,
and by the Russian science support foundation.

\end{document}